\documentclass[aps,pre,onecolumn,superscriptaddress]{revtex4}
\usepackage{graphicx}
\usepackage{color}
\usepackage{amsmath}
\usepackage{amssymb}

\begin{document}

\title{Scaled Brownian motion with renewal resetting}
\author{Anna S. Bodrova}
\address{Humboldt University, Department of Physics, Newtonstrasse 15, 12489 Berlin, Germany}
\author{Aleksei V. Chechkin}
\address{Institute of Physics and Astronomy, University of Potsdam, 14476 Potsdam, Germany}
\address{Akhiezer Institute for Theoretical Physics, Kharkov Institute of Physics and Technology, Kharkov 61108, Ukraine}
\author{Igor M. Sokolov}
\address{Humboldt University, Department of Physics, Newtonstrasse 15, 12489 Berlin, Germany}

\begin{abstract}
We investigate an intermittent stochastic process in which the diffusive motion with time-dependent diffusion coefficient
$D(t) \sim t^{\alpha -1}$ with $\alpha > 0$ (scaled Brownian motion) is stochastically reset to its initial position, and starts anew. 
\color{black}
In the present work we discuss the situation, in which the memory on the value of the diffusion coefficient at a resetting time 
is erased, so that the whole process is a fully renewal one. The situation when the resetting of coordinate does not affect the diffusion 
coefficient's time dependence is considered in the other work of this series. We show that the properties of the probability densities in such processes
(erazing or retaining the memory on the diffusion coefficient) are vastly different. \color{black} In addition we discuss the first passage properties
of the scaled Brownian motion with renewal resetting and consider the dependence of the efficiency of search on the parameters of the process. 
\end{abstract}

\maketitle

\section{Introduction}

\color{black}
Resetting represents a class of stochastic processes, when a random motion is from time to time terminated and
restarted from given initial conditions. The random motion under stochastic resetting arises as the interplay of two distinct random processes: the
resetting process, a point process on the time axis, and particle's motion between the resetting events, which we will call
the displacement process. 

The present work belongs to a series of two works where we consider random processes arising from the 
resetting of a scaled Brownian motion (SBM) to its initial position. Scaled Brownian motion is the diffusion process with explicitly 
time-dependent diffusion coefficient $D(t)\sim t^{\alpha-1}$. Therefore when discussing resetting of the scaled Brownian motion two distinct situations
can be considered. The first one corresponds to the case when after the resetting at time $r_r$ the SBM process starts anew, i.e. with  $D(t)\sim (t-t_r)^{\alpha-1}$, 
so that all motions within the epochs between two subsequent resetting events can be considered as statistical copies of each other. The overall process is then a renewal one. 
The second one is pertinent to the case when resetting of the coordinate does not  affect the 
explicit time-dependence of the diffusion coefficient. This corresponds to retaining of the memory on the instant of time when the SBM started, and
motions within the epochs between two subsequent resetting events are statistically different. 
The difference between the frist, renewal, and the second, non-renewal, situation (full vs. partial resetting) is of importance for all displacement processes 
with non-stationary increments, and was stated e.g. in \cite{Sh2017}, where a
continuous time random walk (CTRW) model under renewal resetting was discussed. However, we are not aware of any work where the two situations were compared for the same 
displacement process to stress and quantify the differences between them. Considering SBM delivers a unique opportunity to make such a comparison:  
the simplicity of the displacement process allows for detailed analytical discussion of the properties of the probability density functions (PDFs) of SBM under resetting in both situations.

There are several reasons to discuss the problems of the fully renewal and non-renewal SBM-resetting process in two separate papers.
First, the two situations require slightly different mathematical approaches, and second, since the renewal situation is much simpler, 
a larger program of investigations may be performed for this case. Moreover, the renewal situations are the ones typically considered in other
processes, so that the results can be immediately compared to such. Notably,
the renewal nature of the process allows for a very simple discussion of its first passage properties \cite{SokChech}, a problem which
is not considered in the parallel publication dealing with the non-renewal case \cite{Anna0}. We note that a very general approach to deal with the resetting problem in a renewal 
setting was given very recently in \cite{Maso1}. The authors systematically studied the effect of Poissonian and power law resetting on an underlying process which, in the absence of resetting, 
displays anomalous diffusion properties, $\langle x^2(t) \rangle \propto t^\alpha$, $0 < \alpha < 2$. In the present work we consider a broader domain of parameters 
for a particular model of transport process, and show additional interesting regimes that appear in the domains not discussed in \cite{Maso1}.
\color{black}

The interest to the first passage properties of stochastic processes under resetting was motivated by the problems of optimal search \cite{Search}.
Since searching for a target at an unknown location may go in a completely wrong direction, it might be useful to return to the initial point, and to start
from the very beginning. This behavior can be modeled by a stochastic exploration process, which is interrupted by random resettings to the initial
position. Examples of such processes are found in many fields such as computer science \cite{computerscience}, biochemistry \cite{chemistry} and \textcolor{black}{biology \cite{biology,bio1,bio2,bio3}}. 
In computer science random walks with stochastic restarts represent a useful strategy to optimize search algorithms in hard
computational problems \cite{computerscience}. In biochemistry, resetting, connected with a departure of an enzyme
from a substrate, may increase the rate of product formation  \cite{chemistry}. 
In biology, gene transcription by RNA polymerases can be viewed as a diffusion process with
stochastic resetting  \cite{biology}. 

One of the main characteristics of the efficiency of a search process is the mean first passage time (MFPT) - the time, at which the particle hits
the target for the first time on the average \cite{redner, redner2}.
Early work concentrated on a searcher performing one-dimensional Brownian motion with Poissonian resetting \cite{EvansMajumdar,EM2011,evma13}. While the 
MFPT to a target for a diffusing particle in absence of resetting diverges, in presence of resetting it turns finite, and there exist an optimal rate of resetting which minimizes 
the MFPT. The discussion has been extended to two and higher dimensions in \cite{high12,bhat}. While for ordinary
Brownian motion the optimal resetting rate depends only on the distance $x_0$ between the target and the origin, and on the diffusion coefficient $D$ of the particle, 
in the case of active particles it may depend on finer details of motion not comprised in the effective diffusion coefficient \cite{active}. 
Another search process is the one in which an individual searcher has a random, finite lifetime, and, when it expires, is replaced by 
a new one starting at the origin \cite{gelenbe}. 
\textcolor{black}{Usually, one considers situations when resetting to the initial position takes place instantaneously and the next motion epoch
follows immediately. This model does not closely model e.g. motion of foraging animals, where introducing refractory periods after return may provide a better description.
Such reset-and-wait models were considered in \cite{wait2, wait3}.}

Refs. \cite{Levy1,Levy2} discuss the MFPT for the case of L\'evy flights with discrete time and Poissonian resetting, respectively,
and obtain the parameter values minimizing the MFPT for several cases including random distribution of targets. 
Other types of the waiting time distributions for resetting events have been studies in \cite{palrt,res2016,shlomi2016,shlomi2017}. It has been shown that the resetting at a constant pace 
is the most effective strategy \cite{SokChech,bhat,palrt,shlomi2017}. 
However, physical constraints on the restart process (e.g. unavoidable stochastic fluctuations in biological systems) make implementation of this optimal protocol inviable \cite{krishna}.
Ordinary Brownian motion with power law waiting time densities for resetting events has been investigated in \cite{NagarGupta}. It has been shown
that there exist a certain power-law exponent minimizing the MFPT for a given target position.

In the present work we investigate particles performing scaled Brownian motion (SBM) \cite{LimSBM,MetzlerSBM,SokolovSBM}, a stochastic process with
time-dependent diffusion coefficient, between the resetting events. We calculate the mean-squared displacement (MSD), the probability density function (PDF) of this
process and the MFPT to a given target. In the other paper of this series \cite{Anna0} we have consider the non-renewal process when the diffusion coefficient
remains unaffected by the resetting events. In the present work we concentrate on the renewal situation, where the diffusion coefficient is also
reset, and the displacement process is rejuvenated under resetting so that the whole process is a renewal one. For this process a larger program of investigation than in
\cite{Anna0} can be performed. The plan of the paper is as follows. In the next Section II we give a brief overview of the properties of SBM. 
In Section III we introduce the main quantities used in the resetting theory, provide general analytic expressions for MSD and PDF and describe the algorithm of numerical simulations. In Section IV 
we calculate MSD and PDF for Poissonian resetting, and in Section V - the MFPT. In Sections VI and VII we discuss power-law resetting 
for $0<\beta<1$ and $\beta>1$, correspondingly, and in Section VIII derive first-passage time for power-law resetting. We give our conclusions in Section IX.

\section{Scaled Brownian motion}

\textcolor{black}{As already stated in the Introduction,} scaled Brownian motion \cite{LimSBM} is the diffusion process with explicitly time-dependent diffusion coefficient $D(t)\sim t^{\alpha-1}$, in
which the mean squared displacement grows as
\begin{equation}\left< x^2(t) \right> = 2K_{\alpha}t^{\alpha}
\label{x2-AD}\end{equation}
with $K_\alpha$ being the generalized diffusion coefficient. For $\alpha=1$ the process is identical to normal, Fickian diffusion, for $\alpha > 1$ 
it is super- and for $0 < \alpha < 1$ subdiffusive. We define SBM in terms of the stochastic process
\begin{equation}
\frac{dx(t)}{dt}=\sqrt{2D(t)}\eta(t)
\end{equation}
with Gaussian noise $\eta(t)$ possessing zero mean $\left\langle \eta(t) \right\rangle = 0$ and the correlation function
\begin{equation}
\left\langle \eta(t_1)\eta(t_2) \right\rangle = \delta\left(t_1-t_2\right)\,.
\end{equation}
The time-diffusion coefficient has the following form:
\begin{equation}
D(t)= \alpha K_{\alpha}t^{\alpha-1}.
\label{Dt}
\end{equation}
The probability density function (PDF) of displacements of particles performing free SBM starting at time $t'$ is Gaussian,
\begin{equation}
p_0\left(x,t-t^{\prime}\right)=\frac{1}{\sqrt{4\pi
K_{\alpha}\left(t-t^{\prime}\right)^{\alpha}}}\exp\left(-\frac{x^2}{4K_{\alpha}\left(t-t^{\prime}\right)^{\alpha}}\right).
\label{p0sbm}
\end{equation}

\section{SBM with stochastic resetting: General expressions}

\color{black}

Let us assume that the particle starts its motion at the origin and returns there at resetting events. The simplest and most studied case corresponds to exponentially distributed waiting times between resets,
\begin{equation}
\psi (t) = r{e^{ - rt}}\,.
\label{pdfexp}
\end{equation}
We also consider the power-law distribution of the waiting times
\begin{equation}
\psi (t) = \frac{\beta/\tau_0}{\left(1+t/\tau_0\right)^{1+\beta}}\,,
\label{pdfpow}
\end{equation}
where $\tau_0$ is a constant. The probability that no resetting events occur between $0$ and $t$ (the survival probability) can be expressed as
\begin{equation}
\Psi (t) = 1 - \int\limits_0^t {\psi (t')dt'}\,.
\label{surv}
\end{equation}
For exponential resetting this is given by
\begin{equation}
\Psi (t) =  {e^{ - rt}}\,.
\label{Psit}
\end{equation}
and for the power-law resetting by 
\begin{equation}
\Psi (t) = \left(1+t/\tau_0\right)^{-\beta}.
\label{psireset}
\end{equation}
The rate of resetting events 
\begin{equation}
\kappa(t)=\sum_{n=1}^{\infty}\psi_n(t),
\end{equation}
where $\psi_n(t)$ is the probability density of the time of $n$-th resetting event. These densities satisfy the recursion relation 
\begin{equation}
\psi_n(t) = \int_0^{t}\psi_{n-1}(t^{\prime})\psi(t-t^{\prime})dt^{\prime}. 
\end{equation}
In the Laplace domain we then get
\begin{equation}
 \tilde{\psi}_n(s) = \tilde{\psi}_{n-1}(s) \tilde{\psi}(s) = \tilde{\psi}^n(s)
\end{equation}
and
\begin{equation}
\tilde \kappa (s) = \sum\limits_{n = 1}^\infty {{{\tilde \psi }^n}(s)} = \frac{{\tilde \psi (s)}}{{1 - \tilde
\psi (s)}}\,.
\label{phis}
\end{equation}
The case of exponential waiting time distribution corresponds to the resetting events constituting a Poisson process on the line characterized by a constant rate, or intensity, $r$, so that
\begin{equation}
\kappa (t) = r.
\label{phir}
\end{equation}
For the case of the power-law resetting the distinct cases $0 < \beta < 1$ and $1<\beta <2$ and $\beta > 2$ should be considered.
For $\beta>2$ both the first and the second moments of the waiting times PDF do exist:
\begin{eqnarray}
\int_0^{\infty}t\psi(t)dt &=& \frac{\tau_0}{\beta-1}, \\
\int_0^{\infty}t^2\psi(t)dt &=&\frac{2\tau_0^2}{\left(\beta-1\right)\left(\beta-2\right)}.
\end{eqnarray}
For $1<\beta<2$ the second moment does not exist while the first moment does. For $\beta<1$ both the first and
the second moments diverge.

In the Laplace domain
\begin{equation}
\tilde{\psi}(s)= \frac{\beta}{\tau_0}\int_0^{\infty}dte^{-ts}\left(1+\frac{t}{\tau_0}\right)^{-1-\beta}.
\end{equation}
Performing the change of the variables $y=s\left(t+\tau_0\right)$ and integrating by parts we get
\begin{equation}\label{psiss}
\tilde{\psi}(s)=1-e^{s\tau_0}\left(s\tau_0\right)^{\beta}\int_{s\tau_0}^{\infty}dye^{-y}y^{-\beta}. 
\end{equation}
For $s\to 0$ and $0<\beta<1$ the integration yields
\begin{equation}
\label{psib01}\tilde{\psi}(s)=1-\Gamma\left(1-\beta\right)\left(s\tau_0\right)^{\beta}+\ldots\, .
\end{equation}
For $1<\beta<2$ the asymptotic result for $s \to 0$ reads
\begin{equation}
e^{s\tau_0}\int_{s\tau_0}^{\infty}dye^{-y}y^{1-\beta} \to \Gamma\left(2-\beta\right),
\end{equation}
and we get
\begin{equation}
\label{psib12}\tilde{\psi}(s)=1-\frac{s\tau_0}{\beta-1}+\frac{\left(s\tau_0\right)^{\beta}\Gamma\left(2-\beta\right)}{\beta-1}
+ \ldots \, ,
\end{equation}
while for $\beta>2$ we get
\begin{equation}
\label{psib2}
\tilde{\psi}(s)=1-\frac{s\tau_0}{\beta-1}+\frac{\left(s\tau_0\right)^{2}}{\left(\beta-1\right)\left(\beta-2\right)}+\ldots
\, .
\end{equation}
Introducing these expressions for $\tilde{\psi}(s)$ into Eq.(\ref{phis}) and performing the inverse Laplace transform we get the forms of $\kappa(t)$.
Thus, for power-law resetting with $0<\beta<1$ 
\begin{equation}
\kappa(t)=\frac{t^{\beta-1}\tau_0^{-\beta}}{\Gamma(\beta)\Gamma(1-\beta)} .
\label{phib01}
\end{equation}
For power law resetting with $\beta>1$ the rate of resetting events attains a constant value
\begin{equation}
\kappa(t)=\kappa = \frac{\beta-1}{\tau_0}.
\label{phig1}
\end{equation}

The probability to find the particle at location $x$ at time $t$ can be expressed in the following way:
\begin{equation}
p(x,t) = \Psi (t){p_0}(x,t) + \int\limits_0^t dt^{\prime} \kappa (t^{\prime})\Psi (t - t^{\prime})p_0(x,t-t^{\prime}).
\label{eqprob}
\end{equation}
The first term in the right-hand side corresponds to trajectories with no resets, and the second one
accounts for the cases where the last reset took place at time $t'$.

\color{black}

At long times $t\to\infty$ the first term may be safely neglected and the PDF is given only by the second term in Eq. (\ref{eqprob}):
\begin{equation}
p(x,t) \simeq \int\limits_0^t dt^{\prime} \kappa (t^{\prime})\Psi (t - t^{\prime})p_0(x,t-t^{\prime}).
\label{eqprob1}
\end{equation}

Multiplying Eq. (\ref{eqprob}) by $x^2$ and performing the integration with respect to $x$, we get the equation for the MSD
\begin{equation}
\left\langle x^2(t)\right\rangle = 2K_{\alpha} t^{\alpha} \Psi(t) + 2K_{\alpha}\int_0^t dt^{\prime}
\kappa(t^{\prime})\Psi(t-t^{\prime})\left(t-t^{\prime}\right)^{\alpha} .
\label{x2}
\end{equation} 
At long times $t\to\infty$ the first term may be neglected, and we obtain for the MSD
\begin{equation}
\left\langle x^2(t)\right\rangle \simeq 2K_{\alpha}\int_0^t dt^{\prime} \kappa(t^{\prime})\Psi(t-t^{\prime})\left(t-t^{\prime}\right)^\alpha \,.
\label{x21}
\end{equation} 
Below we will investigate the behavior of the PDF, Eq.(\ref{eqprob}) and of the MSD, Eq.(\ref{x2}) for exponential and power-law waiting time PDFs in
the resetting process. Moreover, we will study the first passage properties of the corresponding processes to some point
$x_0$ on a line.

Note that the PDF and the MSD of the process depend only on the Gaussian property of the displacement process at a single time, and therefore will be the same 
for all (Markovian or non-Markovian) Gaussian processes with the same $\langle x^2(t) \rangle$ in free motion, e.g. for the fractional Brownian motion \cite{Schehr},
provided resetting erases memory of the displacement process, and the particle's displacements between the two subsequent resetting events can be considered as 
statistical copies of each other.  

To investigate the first passage time we use the approach put forward in \cite{SokChech}. There, using the renewal property of the whole process, one derives the expression for the PDF of hitting the single target under resetting,
\begin{eqnarray}
\textcolor{black}{\rho(t)} = && \Psi(t) \phi(t) + \int_0^t \psi(t') \Phi(t') \Psi(t-t') \phi(t-t') dt'\\
&& + \int_0^t dt' \psi(t') \Phi(t') \int_0^{t-t'} dt'' \psi(t'') \Phi(t'') \Psi(t-t'-t'') \phi(t-t'-t'') + ... \, ,\nonumber
\end{eqnarray}
where the first term is the probability density of hitting the target in the first successful run which was not reset until the hitting time, 
the second term describes the situation in which the first run was not successful and was terminated by resetting at time $t'$, while the second
run is successful, etc. Here $\phi(t)dt$ is the probability that the target is hit in the time interval between $t$ and $t+dt$, and was never hit before,
and $\Phi(t) = 1 - \int_0^t \phi(t') dt'$ is the survival probability of a target in a single run. The cumulative distribution function for hitting in an 
uninterrupted run will be denoted by $F(t) = \int_0^t \phi(t') dt'$.
Denoting $K(t) = \Psi(t) \phi(t)$ and $R(t) = \psi(t) \Phi(t)$, and turning to the Laplace domain, we get
\begin{equation}
 \tilde{\rho}(s) = \frac{\tilde{K}(s)}{1-\tilde{R}(s)} 
 \label{Gen}
\end{equation}
with $\tilde{K}(s)$ being the Laplace transform of $K(t)$, and $\tilde{R}(s)$ is the Laplace transform of $R(t)$. 
The Laplace transform $\tilde{P}(s)$ of the survival probability $P(t)= 1 - \int_0^t \rho(t') dt'$ under resetting is then given by 
\begin{equation}
 \tilde{P}(s) = \frac{1}{s} - \frac{1}{s} \frac{\tilde{K}(s)}{1-\tilde{R}(s)} = \frac{\tilde{X}(s)}{1-\tilde{R}(s)} 
 \label{eq:Q}
\end{equation}
with  $\tilde{X}(s) = s^{-1}[1 - \tilde{R}(s) - \tilde{K}(s)]$. The mean first passage time $\tau = \int_0^\infty P(t)dt$ under resetting follows then in form \cite{SokChech} 
\begin{equation}\label{fpt}
\tau=\frac{X_0}{1-R_0}
\end{equation}
with
\begin{equation}
\label{X0}
X_0=\tilde{X}(0)=\int_0^{\infty}dt \Phi (t) \Psi(t)
\end{equation}
and
\begin{equation}
\label{R0}
R_0=\tilde{R}(0)=\int_0^{\infty}dt \Phi(t)\psi(t). 
\end{equation}
\color{black}
Another approach can be used for calculation of the MFPT \cite{shlomi2017}:
\begin{equation}\label{eqshlomi}
\tau=\frac{\left\langle \min (T,T_r)\right\rangle}{\mbox{Pr} \left(T<T_r\right)}\,,
\end{equation}
where $T$ is a random duration of a single run, $T_r$ is the random time of restart, and $\mbox{Pr}\left(T<T_r\right)$ is the probability that $T<T_r$.
It can be shown that this equation is equivalent to Eq.~(\ref{fpt}) in the following way (see Supplemental Material of Ref. \cite{SokChech}).
The probability that either the reset will take place or the target will be hit is $1-\Psi(t)\Phi(t)$, and the numerators of Eqs. (\ref{fpt}) and (\ref{eqshlomi}) are equal:
\begin{equation}
\left\langle \min(T,T_r)\right\rangle=\int_0^{\infty} t \frac{d}{dt} \left[1-\Psi(t)\Phi(t)\right] dt =\int_0^{\infty} \Phi (t) \Psi(t) dt =\tilde{X}_0\,.
\end{equation}
The denominators are also equal according to:
\begin{equation}
1-\tilde{R}_0=\int_0^{\infty} \left[ 1-\Phi(t) \right] \psi(t)dt = \int_0^{\infty}F(t)\frac{d\Psi(t)}{dt}dt =\int_0^{\infty} \Psi(t)\frac{dF(t)}{dt} dt =\int_0^{\infty}\Psi(t)\phi(t) dt
\end{equation}
which is nothing else that probability that $T<T_r$. Here $F(t)=1-\Phi(t)$ is the total hitting probability.
\color{black}

Within this work we will compare analytical results with numerical simulations. In the renewal resetting the whole process starts anew at the resetting event, 
and the memory on its previous course is erased. Therefore the
simulation of the process up to the last resetting event is not necessary to get the MSD and the PDF of the particle's positions. 
The event-driven simulations for MSD and for PDF are performed as follows. For a given sequence of the output times $t$ we simulate the sequence of resetting events, find the time of
the last resetting event $t' < t$ and set $x(t')=0$. Then the position of the particle at time $t$ is distributed according to a Gaussian
with zero mean and variance $\langle x^2 (t) \rangle = 2 K_\alpha (t - t')^\alpha$. The corresponding Gaussian can be obtained from 
a standard normal distribution generated using the Box-Muller transform. The results are averaged over $N=10^4$ to $10^6$ independent runs.

In order to calculate the first passage time we apply the direct simulation. The time axis is discretized with the step $dt=t_{i+1}-t_i$, and the time
of the first resetting $T_1$ is generated according to the waiting time density $\psi(t)$. Then the particle's motion is simulated, and, if the 
target was not hit up to $T_n$, the coordinate is reset to $x=0$, a new resetting time $T_2$ is generated, and the simulation repeated, etc.
The particle's motion between the resetting events is modeled by a finite-difference analogue of the Langevin equation
\begin{eqnarray}\label{lattice}
x_{i+1}=x_i+\xi_i\sqrt{2\alpha K_{\alpha}(t_i-T_n)^{\alpha-1}dt}
\end{eqnarray}
(for $T_n < t_i < T_{n+1}$).  Here $x_i=x(t_i)$ is the coordinate of the particle at the time $t_i$, and
$\xi_i$ is the random number distributed according to a standard normal distribution generated using the Box-Muller transform.
The simulation stops when the particle's coordinate exceeds $x_0$ for the first time.
The first hitting time is then averaged over $N=10^4$ to $10^5$ realizations. The value of $\alpha K_\alpha$ is set to unity in all our simulations.
In simulations corresponding to power-law waiting time distributions we set $\tau_0=1$. 

\section{MSD and MFPT for Poissonian resetting}

\subsubsection{Mean squared displacement} 
\begin{figure}[htbp]
  \centerline{
\includegraphics[width=0.7\columnwidth]{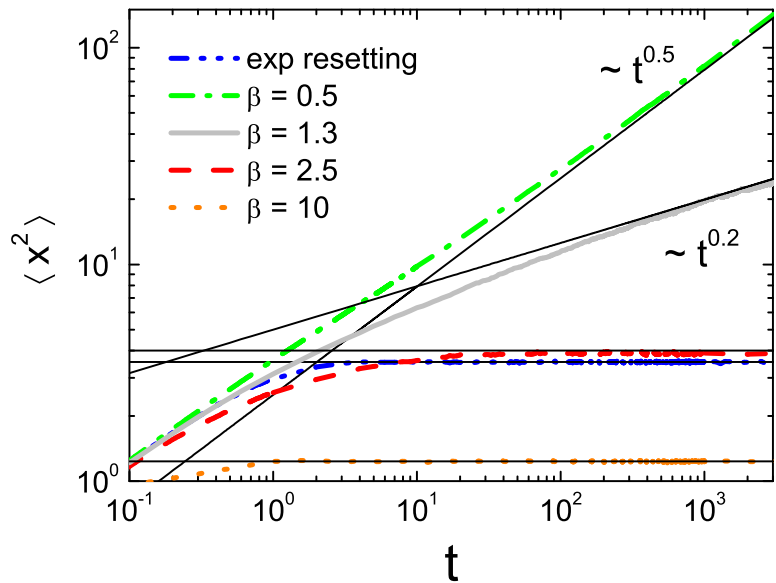}}
\caption{Mean-squared displacement for power-law and Poissonian resetting with $r=1$ and for power law resetting with $\alpha=0.5$ as obtained in $N=10^4$ realizations. Results of simulation are shown by thick colored lines 
and the analytical asymptotics by thin black solid lines (Eq.~(\ref{x2exp}) for exponential resetting, Eq.~(\ref{msd0}) for power-law resetting with \textcolor{black}{$\beta=0.5$, Eq.~(\ref{bla1}) 
for $\beta=1.3$, Eq.~(\ref{ba1}) for $\beta=2.5$ and 10}).}
\label{GR2exprenewal}
\end{figure}

The MSD for SBM with exponential resetting can be obtained by inserting Eqs. (\ref{Psit}) and (\ref{phir}) into Eq. (\ref{x2}),
and reads
\begin{equation}
\left\langle x^2(t)\right\rangle = 2K_{\alpha}t^{\alpha}e^{-rt}+\frac{2K_{\alpha}}{r^{\alpha}}\gamma\left(\alpha+1,rt\right)
\end{equation}
with $\gamma(a,z)$ being the lower incomplete Gamma-function
\begin{equation}
\gamma\left(a,z\right)=\int_0^{z}dx e^{-x}x^{a-1}.
\end{equation}
Expanding the incomplete Gamma-function for $z\to\infty$
\begin{equation}
\gamma\left(a,z\right) \simeq \Gamma(a)-z^{a-1}e^{-z}-\frac{a-1}{z}z^{a-1}e^{-z},
\end{equation}
we obtain for the MSD
\begin{equation}
\left\langle x^2(t)\right\rangle \simeq \frac{2K_{\alpha}}{r^{\alpha}} \left[\Gamma\left(\alpha+1\right)-\alpha\left(rt\right)^{\alpha-1}e^{-rt}\right].
\end{equation}
It rapidly tends to the steady state:
\begin{equation}\label{x2exp}
\left\langle x^2(t)\right\rangle = \frac{2K_{\alpha}}{r^{\alpha}}\Gamma\left(\alpha+1\right).
\end{equation}
In Fig.~\ref{GR2exprenewal} we show the analytical result, Eq.~(\ref{x2exp}), together with the results of direct numerical simulation (blue line), and get an
excellent agreement.  \textcolor{black}{Fig. \ref{GR2exprenewal} gives the overview of all MSD behaviors discussed in the present paper: Other lines in Fig.~\ref{GR2exprenewal} show the results 
for the MSD in the cases of power-law resetting considered in Sections 6 and 7. The Figure will be referred to several time throughout the text. }

For $\alpha=1$ we recall the ordinary Brownian motion between the resetting events and get at long times the stationary value of MSD
\begin{equation}
\left\langle x^2(t)\right\rangle = \frac{2 K_{1}}{r}\,.
\label{x2expobm}
\end{equation}

\subsubsection{Probability density function} 
The probability density function for Poissonian resetting may be obtained by inserting Eqs. (\ref{p0sbm}), (\ref{Psit})
and (\ref{phir}) into Eq. (\ref{eqprob1}) which for long times yields
\begin{equation}
p\left(x,t\right)\simeq r\int_0^tdt^{\prime}e^{-r\left(t-t^{\prime}\right)}\frac{1}{\sqrt{4\pi
K_{\alpha}\left(t-t^{\prime}\right)^{\alpha}}}\exp\left(-\frac{x^2}{4K_{\alpha}\left(t-t^{\prime}\right)^{\alpha}}\right).
\label{eq:26}
\end{equation}
Introducing a new variable $\tau=t-t^{\prime}$ we rewrite the integral as 
\begin{equation}\label{strangeint}
p\left(x,t\right) \simeq \frac{r}{\sqrt{4\pi K_{\alpha}}}\int_0^t d\tau\exp\left(-\varphi(\tau)\right)\tau^{-\frac{\alpha}{2}}
\end{equation}
with the function $\varphi(\tau)=r\tau+\frac{x^2}{4K_{\alpha}\tau^{\alpha}}$ which attains a simple quadratic minimum at 
\begin{equation}
\tau_{\min} = \left(\frac{\alpha x^2}{4K_{\alpha}r}\right)^{\frac{1}{\alpha+1}}.
\end{equation}
For $0 \ll \tau_{\min} \ll t$ the standard Laplace method can be used:
\begin{equation}
p\left(x,t\right) \simeq \frac{r}{\sqrt{4\pi K_{\alpha}\tau_{\min}^{\alpha}}}\exp\left(-\varphi(\tau_{\min})\right)\int_{-\infty}^{\infty}
\exp\left(-\frac12\varphi^{\prime\prime}(\tau_{\min}) \left(\tau-\tau_{\min}\right)^2\right) d\tau,
\end{equation}
provided $\varphi^{\prime\prime}(\tau_{\min})$ is large enough. Both conditions correspond to the intermediate asymptotic behavior in $x$.  The final result for this case 
corresponds to a stationary state and reads:
\begin{equation}
p(x) \simeq \frac{r\sqrt{2}}{\sqrt{\alpha\left(\alpha+1\right)}}\left(\frac{\alpha}{4K_{\alpha}r}\right)^{\frac{1}{\alpha+1}} |x|^{\frac{1-\alpha}{\alpha+1}}\exp\left(-\left(\frac{x^2
r^{\alpha}}{4K_{\alpha}}\right)^{\frac{1}{\alpha+1}}\left(\alpha^{\frac{1}{\alpha+1}}+\alpha^{-\frac{\alpha}{\alpha+1}}\right)\right). \label{pdfexp1}
\end{equation}

\begin{figure}[htbp]
  \centerline{
\includegraphics[width=0.7\columnwidth]{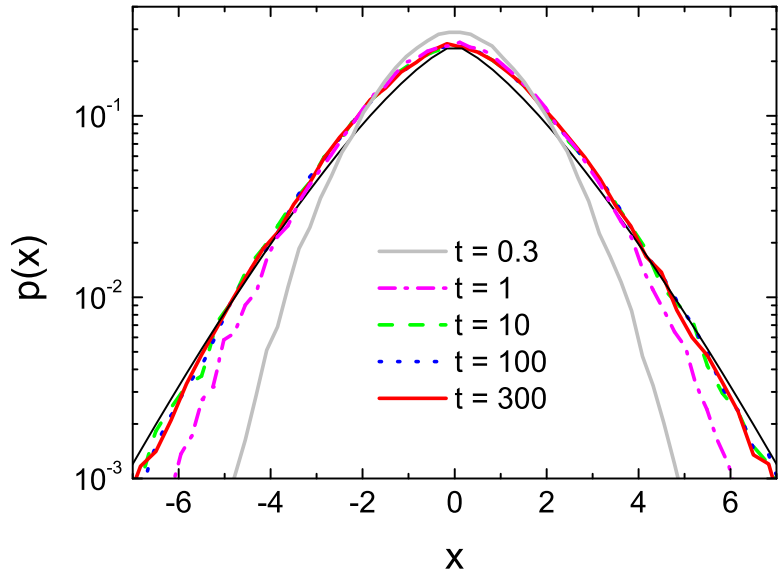}}
\caption{PDF for SBM with $\alpha=0.5$ and with Poissonian resetting at times $t=0.3, 1, 10, 100, 300$ showing the convergence to a steady state. The theoretical prediction is shown by the thin black solid line, see text for details.
Parameters: $N=10^5$, $r=1$.}
\label{GSBMexprenewal}
\end{figure}

Fig.~\ref{GSBMexprenewal} presents  the results of numerical simulation for $\alpha = 0.5$. The theoretical curve (thin black line) corresponds to the exponential term of Eq.(\ref{pdfexp1}) 
with the pre-exponential factor replaced by the actual value of $p(x)$ for $x=0$ calculated from Eq.(\ref{eq:26}). 

For $\alpha=1$, the PDF behaves as that for ordinary Brownian motion with exponential resetting, namely, is a 
two-sided exponential  (Laplace) distribution as obtained in \cite{EvansMajumdar, EM2011}:
\begin{equation}
p\left(x\right)=\frac12\sqrt{\frac{r}{K_1}}\exp\left(-\sqrt{\frac{r}{K_1}}\left|x\right|\right).
\end{equation}

\section{MFPT for Poissonian resetting}

Now let us calculate the mean time needed for the particle which starts at $x=0$ to hit a target at position $x_0 \neq 0$. The hitting time probability in a single run of
the SBM is given by a change of variables in the L\'evy-Smirnov distribution and reads 
\begin{equation}
\phi (t)=\frac{\alpha x_0}{\sqrt{4\pi K_{\alpha}}}\exp\left(-\frac{x_0^2}{4K_{\alpha}t^{\alpha}}\right)t^{-1-\frac{\alpha}{2}} .
\end{equation}
The survival probability of a target in a single run may be obtained via the integration of the hitting time probability over $t$ 
\begin{equation}\label{Phit}
\Phi (t)=1-\int_0^t \phi (t^{\prime})dt^{\prime} =\mathrm{erf}\left(\frac{x_0}{2\sqrt{K_{\alpha}t^{\alpha}}}\right) .
\end{equation}
Taking into account that for exponential resetting $\psi(t)=r\Psi(t)$ as given by Eqs. (\ref{pdfexp}) and (\ref{Psit}), one gets $R_0=rX_0$, and the MFPT as given by Eq.(\ref{fpt}) can be represented as
\begin{equation}
\label{fpt1}
\tau=\frac{X_0}{1-rX_0} .
\end{equation}

The expression for $X_0$ follows by inserting Eq.~(\ref{pdfexp}), Eq.~(\ref{Psit}) and Eq.~(\ref{Phit}) into Eq.~(\ref{X0}): 
\begin{equation}
X_0=\frac{2}{\sqrt{\pi}}\int_0^{\infty}dt e^{-rt}\int_0^{\frac{x_0}{2\sqrt{K_{\alpha}t^{\alpha}}}}da e^{-a^2}.
\end{equation}
This expression can be rewritten by changing the order of integrations
\begin{equation}
X_0=\frac{2}{\sqrt{\pi}}\int_0^{\infty}da e^{-a^2} \int_0^{\left(\frac{x_0^2}{4 K_{\alpha}a^{2}}\right)^{1/\alpha}}e^{-rt} dt ,
\end{equation}
after which the inner integral can be evaluated explicitly: 
\begin{equation}
X_0=\frac{1}{r}-\frac{2}{r\sqrt{\pi}}\int_0^{\infty}da\exp\left(-a^2-r\left(\frac{x_0^2}{4K_{\alpha}a^2}\right)^{\frac{1}{\alpha}}\right) .
\end{equation}
The integral can be estimated by using the Laplace method which leads to
\begin{equation}\label{X00}
X_0 \simeq \frac{1}{r}\left(1-\sqrt{\frac{2\alpha}{\alpha+1}}\exp\left(-\left(\alpha^{\frac{1}{1+\alpha}}+\alpha^{-\frac{\alpha}{1+\alpha}}\right)r^{\frac{\alpha}{\alpha+1}}\left(\frac{x_0^2}{4K_{\alpha}}\right)^{\frac{1}{\alpha+1}}\right)\right) .
\end{equation}
Inserting this into Eq.~(\ref{fpt1}) we arrive at the final expression for mean first passage time
\begin{equation}\label{tauexp}
\tau \simeq \frac{1}{r}\left\{\sqrt{\frac{\alpha+1}{2\alpha}}\exp\left[r^{\frac{\alpha}{\alpha+1}}\left(\frac{x_0^2}{4K_{\alpha}}\right)^{\frac{1}{\alpha+1}}\left(\alpha^{\frac{1}{1+\alpha}}+\alpha^{-\frac{\alpha}{1+\alpha}}\right)\right]-1\right\} .
\end{equation}
For ordinary Brownian diffusion ($\alpha=1$) we get the known expression \cite{EvansMajumdar}, which is now exact:
\begin{equation}
\tau = \frac{1}{r}\left[\exp\left(\left|x_0\right|\sqrt{r/K_{\alpha}}\right)-1\right] .
\end{equation}

{\begin{figure}[htbp]
  \centerline{\includegraphics[width=0.6\columnwidth]{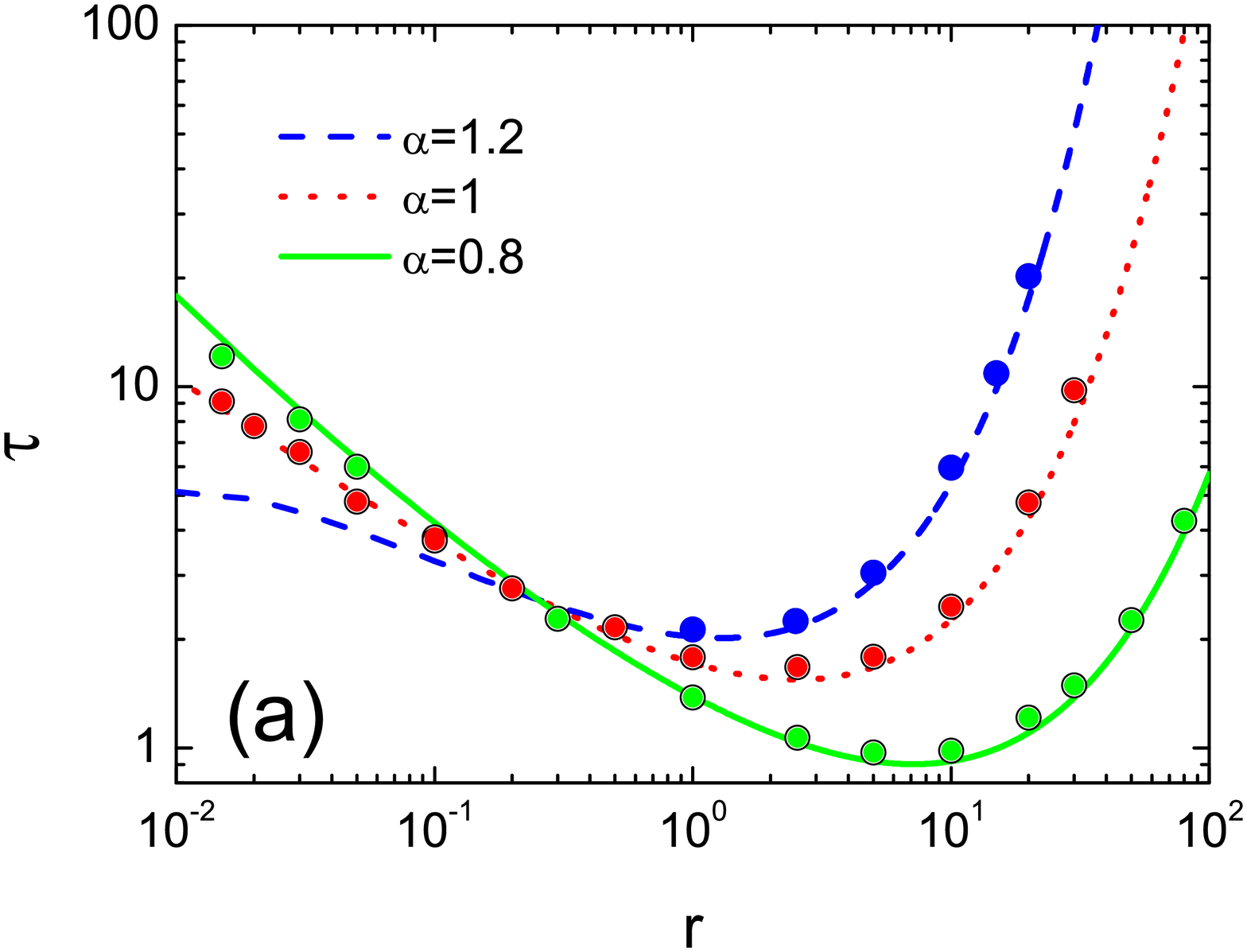}
\includegraphics[width=0.6\columnwidth]{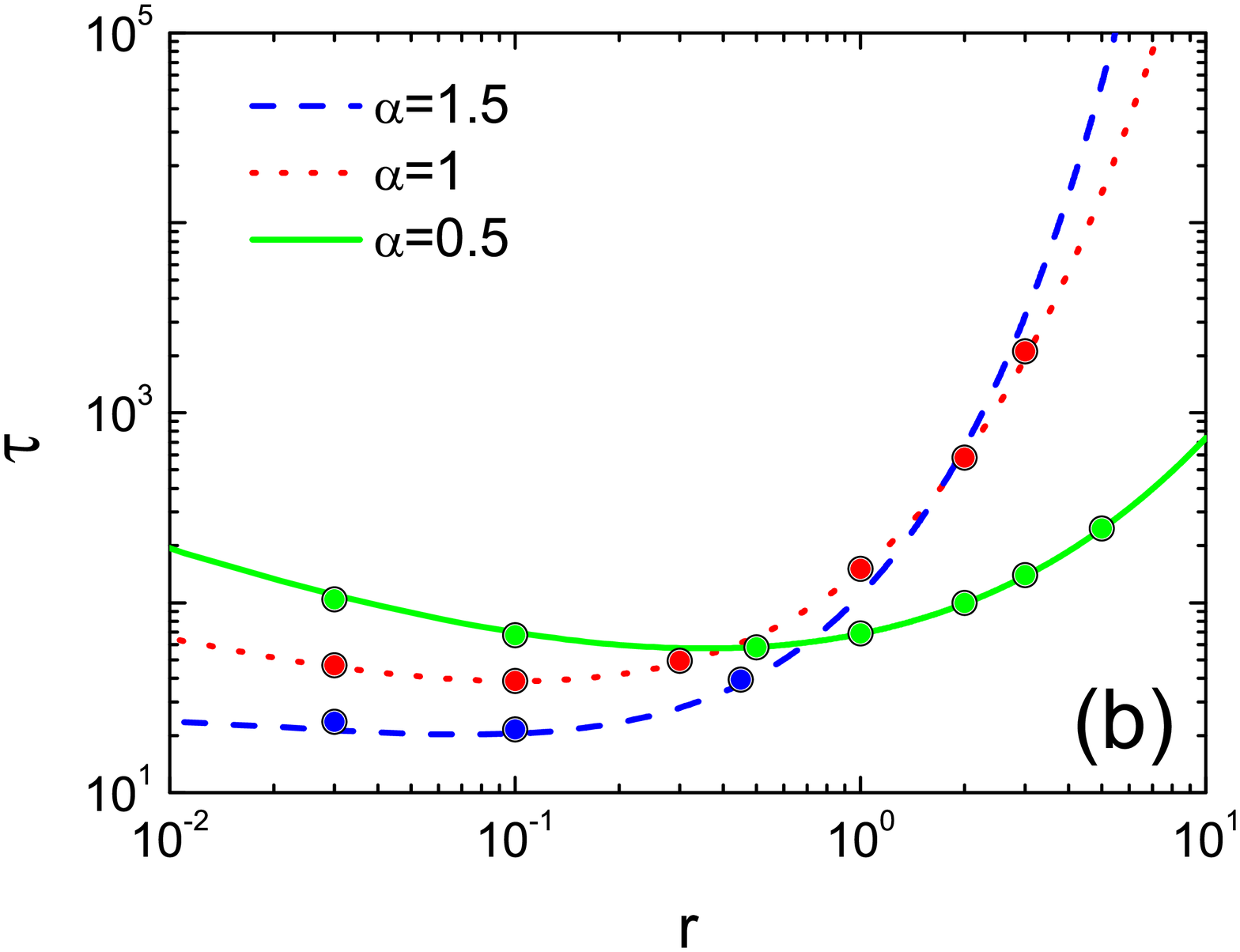}}
\caption{The mean first passage time $\tau$ versus the resetting rate $r$ for different $\alpha$ values. The target positions are \textcolor{black}{ $x_0=1$ (a) and  $x_0=5$ (b)}. Results of simulations are shown by circles and the analytical results are represented by the lines according to Eq.~(\ref{tauexp}).}
\label{GTx5}
\end{figure} }

The MFPT given by Eq.~(\ref{tauexp}) for different parameter values is shown at Fig. \ref{GTx5}. At very low rates $r\to 0$ corresponding
to free scaled Brownian motion without resetting the MFPT tends to infinity. At $r\to\infty$ the MFPT also tends to infinity, because the particle
does not have enough time to locate the target between the resetting events.
Therefore there exists an optimal rate $r^{*}$ which minimizes the MFPT. 
Let us first consider the situation where the target is located relatively close to the initial position of the particle ($x_0=1$), Fig.~\ref{GTx5}a.
In this case the superdiffusive search process is more effective (i.e. leads to shorter MFPT) at relatively small rates $r \leq 0.2$, while at larger rates the target will be
found faster under subdiffusive motion. If the distance between the target and initial position of particle increases ($x_0=5$), the superdiffusion
becomes more favorable for a wider range of $r$ (Fig. \ref{GTx5}b). This resembles search processes with Le\'vy flights where large jumps
are preferable when the target is located far from the initial position and smaller jumps are preferable when the target is
closer to the origin \cite{Palyulin}.

\begin{figure}[htbp]
  \centerline{
\includegraphics[width=0.6\columnwidth]{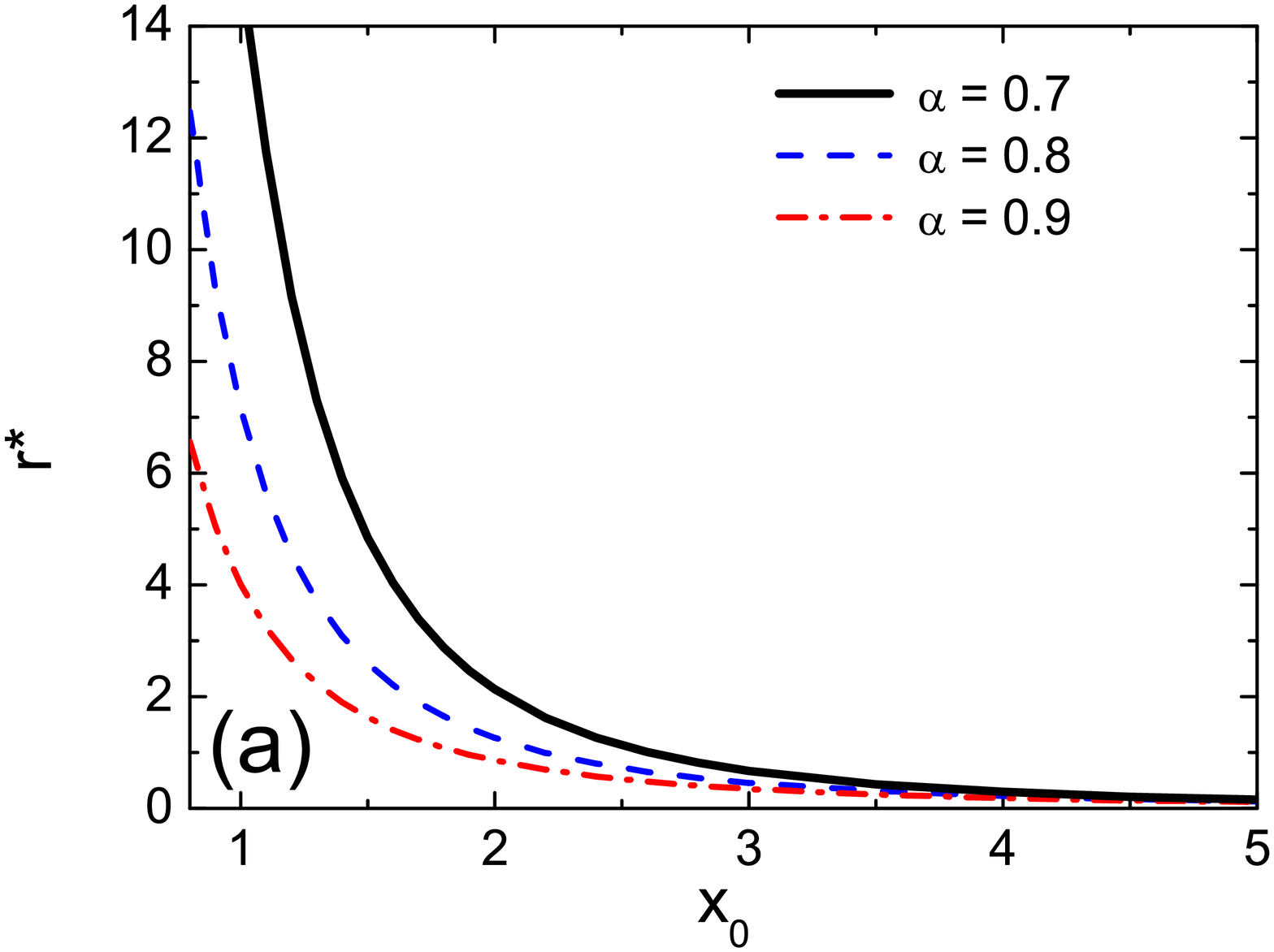}\includegraphics[width=0.6\columnwidth]{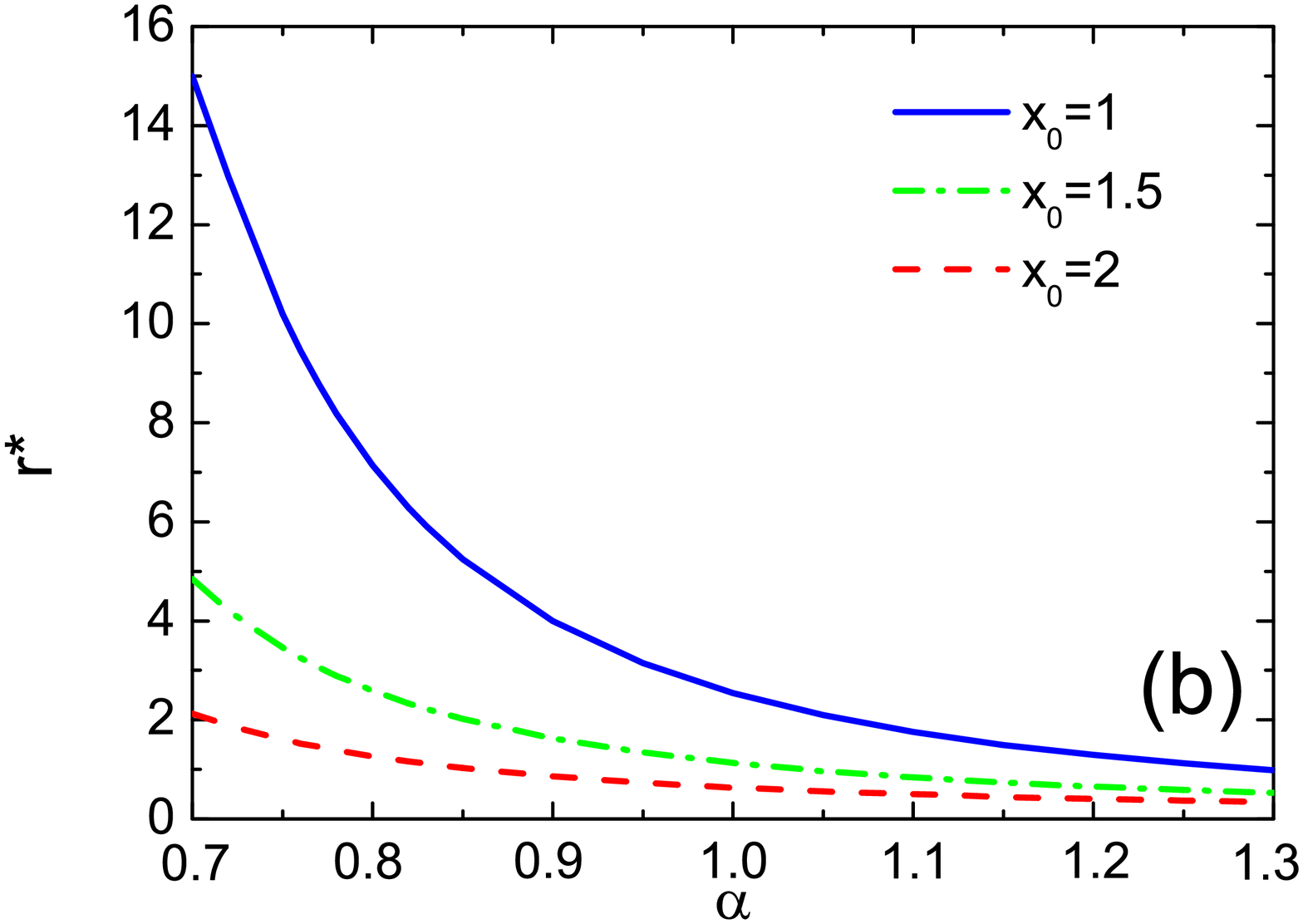}}
\caption{The optimal resetting rate $r^{*}$ (obtained as a numerical solution of Eq.~(\ref{rstar})) \textcolor{black}{ versus the target position $x_0$ for $\alpha=0.7, 0.8, 0.9$ (a) and versus $\alpha$ for $x_0=1, 1.5, 2$ (b).}}
\label{Gra}
\end{figure}

Minimizing $\tau$ (Eq.~\ref{tauexp}) with respect to $r$ we get the equation for the optimal resetting rate $r^*$ 
\begin{equation}\label{rstar}
1-\sqrt{\frac{2\alpha}{\alpha+1}}\exp\left(-B r^{*c}\right)=Bcr^{*c}
\end{equation}
with
\begin{equation}
c = \frac{\alpha}{\alpha+1} 
\end{equation}
and
\begin{equation}
B = \left(\frac{x_0^2}{4K_{\alpha}}\right)^{\frac{1}{\alpha+1}}\left(\alpha^{\frac{1}{1+\alpha}}+\alpha^{-\frac{\alpha}{1+\alpha}}\right) ,
\end{equation}
which can be solved numerically. 
Fig.~\ref{Gra} shows that the optimal resetting rate $r^*$ decreases with increase of both the distance $x_0$ between
target and initial position (a) and the exponent $\alpha$ (b). The monotonous decrease of $r^{*}$ as a function $x_0$ parallels to the case of Le\'vy flights \cite{Levy1} but does not 
show the discontinuity observed in this last case.

\section{MSD and PDF with power-law resetting, $0<\beta<1$}

\subsubsection{Mean squared displacement} 

Substituting Eqs. (\ref{psireset}) and (\ref{phib01}) into Eq. (\ref{x21}) we obtain
\begin{equation}
\left\langle x^2(t)\right\rangle = \frac{2K_{\alpha}\tau_0^{-\beta}}{\Gamma\left(\beta\right)\Gamma\left(1-\beta\right)}\int_0^t
dt^{\prime}t^{\prime \beta-1}\left(t-t^{\prime}\right)^{\alpha}\left(1+\frac{t-t^{\prime}}{\tau_0}\right)^{-\beta} .
\end{equation}
Assuming $t \gg \tau_0$ we can neglect unity in the last bracket. Changing to a new integration variable $\tau=t^{\prime}/t$ we get
\begin{equation}
\left\langle x^2(t)\right\rangle = \frac{2K_{\alpha}}{\Gamma\left(\beta\right)\Gamma\left(1-\beta\right)} t^{\alpha} \int_0^1 d\tau
\tau^{\beta-1}\left(1-\tau\right)^{\alpha-\beta},
\end{equation}
so that the integral reduces to a Beta-function, and the whole expression reads
\begin{equation}
\left\langle x^2(t)\right\rangle =\frac{2K_{\alpha}}{\Gamma\left(\beta\right)\Gamma\left(1-\beta\right)}B\left(\beta,\alpha-\beta+1\right)t^{\alpha}.
\end{equation}
Expressing the Beta-function in terms of Gamma-functions we arrive at the final result
\begin{equation}\label{msd0}
\left\langle x^2(t)\right\rangle
=\frac{2K_{\alpha}\Gamma\left(\alpha-\beta+1\right)}{\Gamma\left(\alpha+1\right)\Gamma\left(1-\beta\right)}t^{\alpha}.
\end{equation}
The time-dependence of MSD remains the same as for the case of free SBM, only the prefactor is altered. 
This result is plotted in Fig.~\ref{GR2exprenewal} and is again in excellent agreement with simulation results given by a green line.
 
For $\alpha=1$ the MSD scales linearly with time
\begin{equation}
\left\langle x^2(t)\right\rangle  = 2K_{1}\left(1-\beta\right)t\,.
\label{x2b01obm}
\end{equation}

\subsubsection{Probability density function} 

The PDF $p(x,t)$ for the case considered shows different behavior for small, intermediate and large values of $x$.
Especially interesting is the case  $\beta > 1-\alpha/2$, when, for $t$ long enough, the PDF develops a pronounced intermediate asymptotic
scaling domain. 

The PDF is obtained by inserting Eqs. (\ref{p0sbm}), (\ref{psireset}) and (\ref{phib01}) into Eq. (\ref{eqprob1}):
\begin{equation}
p(x,t)=\frac{\tau_0^{-\beta}}{\Gamma\left(\beta\right)\Gamma\left(1-\beta\right)}\frac{1}{\sqrt{4\pi K_{\alpha}}}\int_0^t
dt^{\prime}t^{\prime\beta-1}\left(1+\frac{t-t^{\prime}}{\tau_0}\right)^{-\beta}\left(t-t^{\prime}\right)^{-\alpha/2}\exp\left(-\frac{x^2}{4K_{\alpha}\left(t-t^{\prime}\right)^{\alpha}}\right).
\label{IntB}
\end{equation}

To obtain the far tail of the PDF, i.e. its behavior for $|x| \gg \sqrt{4 K_\alpha t^\alpha}$, we change the variable of integration to $y=(1-t'/t)^{-\alpha}$: 
\begin{equation}
p(x,t)=\frac{1}{\alpha\Gamma\left(\beta\right)\Gamma\left(1-\beta\right)\sqrt{4\pi K_{\alpha}t^{\alpha}}}\int_1^{\infty} dy
y^{-\frac{1}{\alpha}-\frac{1}{2}} \left(1-y^{-\frac{1}{\alpha}}\right)^{\beta-1} \left(\frac{\tau_0}{t} + y^{-\frac{1}{\alpha}} \right)^{-\beta} \exp\left(-\frac{x^2y}{4K_{\alpha}t^{\alpha}}\right).
\end{equation}
For $x^2\gg 4K_{\alpha}t^{\alpha}$ the exponential function is decaying very fast, so that the main contribution to the integral stems from the vicinity
of the lower integration bound where we take $y = 1 + \delta$ and approximate in the leading order $ y^{-1 / \alpha-1/2} \simeq 1$,
$(1-y^{-1/\alpha})^{\beta -1} \simeq (\delta/\alpha)^{\beta -1}$ and $\left(\tau_0 / t + y^{-1/\alpha} \right)^{-\beta} \simeq (1+\tau_0/t)^{-\beta} \simeq 1$ assuming $t \gg \tau_0$. We get:
\begin{eqnarray}
 p(x,t) &\simeq& \frac{1}{\alpha\Gamma\left(\beta\right)\Gamma\left(1-\beta\right)\sqrt{4\pi K_{\alpha}t^{\alpha}}}\exp\left(-\frac{x^2}{4K_{\alpha}t^{\alpha}}\right) \alpha^{1 - \beta}
 \int_0^{\infty} d\delta \delta^{\beta-1}\exp\left(-\frac{x^2 \delta}{4K_{\alpha}t^{\alpha}}\right) \nonumber \\
&=& \frac{\alpha^{-\beta}}{\Gamma\left(1-\beta\right)\sqrt{4\pi
K_{\alpha}t^{\alpha}}}\left(\frac{4K_{\alpha}t^{\alpha}}{x^2}\right)^{\beta}\exp\left(-\frac{x^2}{4K_{\alpha}t^{\alpha}}\right).
\label{eqGau}
 \end{eqnarray}
Thus, the behavior of the distribution for  $x^2\gg 4K_{\alpha}t^{\alpha}$, is universally Gaussian, up to a power law correction.

Now we turn to investigation of the PDF's behavior at small and intermediate $|x|$. 
Changing the variable of integration to $a=1-t^{\prime}/t$ we rewrite Eq.(\ref{IntB}) as
\begin{equation}
p(x,t)=\frac{1}{\Gamma\left(\beta\right)\Gamma\left(1-\beta\right)\sqrt{4\pi K_{\alpha}t^{\alpha}}}\int_0^1 da
\left(1-a\right)^{\beta-1}a^{-\alpha/2}\left(\frac{\tau_0}{t}+a\right)^{-\beta}\exp\left(-\frac{x^2}{4K_{\alpha}t^{\alpha}a^{\alpha}}\right),
\label{IntegrA}
\end{equation}
which is the main equation to be analyzed below. The behavior of the distribution for small and intermediate $x$ strongly depends on the relation between $\beta$ and $\alpha$.

Let us first consider the limit $x \to 0$. In this case the exponential in Eq.(\ref{IntegrA}) tends to unity.
For $t\gg\tau_0$ the term $\tau_0/t$ in the last bracket in Eq.(\ref{IntegrA}) may be neglected provided the integral stays convergent under such omission, which is the case for $\beta < 1-\alpha/2$.
For $\beta > 1-\alpha/2$ this term cannot be omitted, since it provides the necessary regularization. As we proceed to show the two cases correspond to vastly different behavior.
In the first case, $0 < \beta < 1 - \alpha/2$ , omitting the corresponding term, we get
\begin{equation}
p(0,t)\simeq\frac{1}{\Gamma\left(\beta\right)\Gamma\left(1-\beta\right)\sqrt{4\pi K_{\alpha}t^{\alpha}}}\int_0^1 da 
\left(1-a\right)^{\beta-1}a^{-\alpha/2-\beta},
\label{III}
\end{equation}
so that 
\begin{equation}
p(0,t)\simeq\frac{1}{\Gamma\left(\beta\right)\Gamma\left(1-\beta\right)\sqrt{4\pi K_{\alpha}t^{\alpha}}} \mathrm{B} \left(-\alpha/2-\beta+1;\beta\right) .
\end{equation}
Expressing Beta-function in terms of $\Gamma$-functions we get an alternative form  
\begin{equation}
p(0,t)=\frac{\Gamma\left(1-\beta-\alpha/2\right)}{\sqrt{4\pi K_{\alpha}t^{\alpha}}\Gamma\left(1-\alpha/2\right)\Gamma\left(1-\beta\right)} .
\label{top1}
\end{equation}
For $\alpha=1$ we obtain, taking into account $\Gamma (1/2)=\sqrt{\pi}$,
\begin{equation}
p(0,t)=\frac{\Gamma\left(\frac12-\beta\right)}{\sqrt{4\pi^2 K_{1}t}\Gamma\left(1-\beta\right)} = \frac{\Gamma\left(1/2-\beta\right)\Gamma\left(\beta\right)}{\sqrt{4\pi K_{1}t}}\frac{\sin (\pi\beta)}{\pi^{3/2}},
\end{equation}
where, in the last expression, Euler's reflection formula $\Gamma\left(1-\beta\right)\Gamma\left(\beta\right)=\frac{\pi}{\sin\pi\beta}$ was used.
This last expression coincides with the result obtained in \cite{NagarGupta} for $\beta<1/2$.

For $\beta>1-\alpha/2$ the integral Eq.(\ref{III}) diverges at $x=0$, and the term $\tau_0/t$ cannot be omitted. The expression for the PDF for  $x=0$ reads
\begin{equation}
p(0,t)=\frac{1}{\Gamma\left(\beta\right)\Gamma\left(1-\beta\right)\sqrt{4\pi K_{\alpha}t^{\alpha}}}\int_0^1 da
\left(1-a\right)^{\beta-1}a^{-\alpha/2}\left(a+\tau_0/t\right)^{-\beta} .
\label{IV}
\end{equation}
The integral can be expressed via a hypergeometric function, see Eq.(2.2.6.15) of \cite{Brychkov}: 
\[
\int_0^1 da \left(1-a\right)^{\beta-1}a^{-\alpha/2}\left(q+a\right)^{-\beta} = q^{-\beta} \mathrm{B} \left(1-\frac{\alpha}{2},\beta \right) \,_2F_1 \left(1-\frac{\alpha}{2},\beta,1-\frac{\alpha}{2}+\beta, -\frac{1}{q} \right).
\]
with $q=\tau_0/t$. For $t \gg \tau_0$ the argument $z = -\frac{1}{q} = -t/\tau_0$ of the hypergeometric function is large and tends to $-\infty$ in the course of time, 
so that $\mathrm{arg}(\beta z) = \pi$, and the asymptotic behavior of this function is given by Eq.(15.7.3) of Ref. \cite{as}. The second term in this expansion 
contains the exponential $e^{\beta z}$ whose argument is negative and large, so that the whole contribution can be neglected. 
Therefore the final approximation in the lowest order reads:
\[
 \;_2 F_1 \left(1-\frac{\alpha}{2},\beta,1-\frac{\alpha}{2}+\beta, -\frac{t}{\tau_0} \right) \simeq \frac{\Gamma\left(1-\frac{\alpha}{2} + \beta \right)}{\Gamma \left( 1 - \frac{\alpha}{2} \right)} \left( \frac{\beta t}{\tau_0} \right)^{\alpha/2 -1}. 
\]
Inserting this approximation into Eq.(\ref{IV}) and expressing the Beta function in terms of Gamma functions we get
\begin{equation}
 p(0,t) \simeq \frac{\Gamma\left(1-\frac{\alpha}{2} + \beta \right)}{\Gamma\left(\beta\right)\Gamma\left(1-\beta\right)\sqrt{4\pi K_{\alpha}\tau_0^{\alpha}}} \left(\frac{t}{\tau_0} \right)^{\beta-1}.
 \label{top}
\end{equation}

\begin{figure}[htbp]
  \centerline{
\includegraphics[width=0.7\columnwidth]{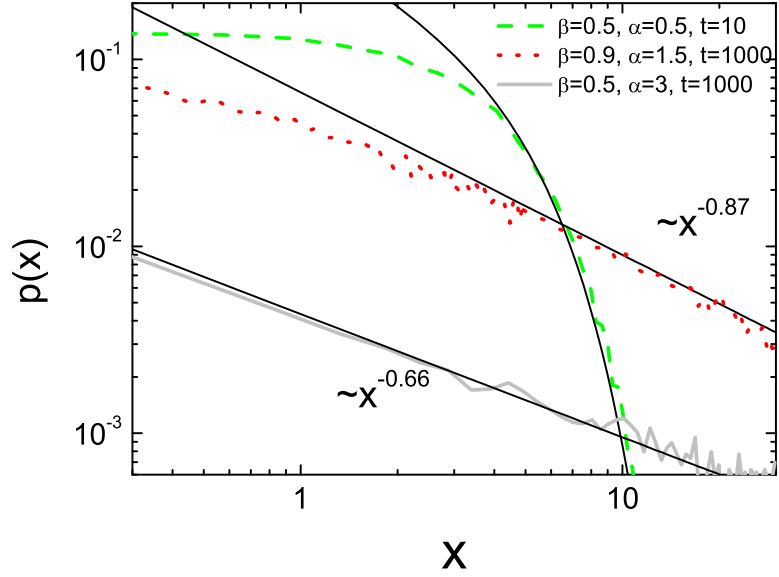}}
\caption{PDF for SBM with power-law resetting with $0<\beta<1$. The dashed line (green online) corresponds to the case $\beta < 1 - \alpha/2$ and shows a flat top of the PDF crossing over to a practically Gaussian behavior, Eq.(\ref{eqGau}), at large $x$. Two other cases correspond to $\beta > 1 - \alpha/2$ and show the emerging
intermediate asymptotic behavior, where $p(x,t) \sim t^{\beta-1}\left|x\right|^{-1-\frac{2\beta}{\alpha}+\frac{2}{\alpha}}$ according to (Eq.~\ref{epdf0b1}) for $\alpha = 1.5$
(red dotted line) and for $\alpha=3$ (gray solid line).
The analytical predictions for the slopes in intermediate asymptotic domains, Eq.(\ref{inter1}), are shown with thin black straight lines. }
\label{Grenewalb05}
\end{figure}

In both cases the PDF does not diverge for $x \to 0$. Now let us consider intermediate values of $|x| \ll \sqrt{4 K_\alpha t^\alpha}$. For $\beta < 1-\alpha/2$ the integral in Eq. (\ref{IntegrA}) converges even in the absence of exponential term, and this term is close to unity in the most part 
of the integration domain except for the vicinity of the lower bound of integration. Therefore the dependence on $x$ is weak, and the PDF at small $x$ has a flat top where its value is given by Eq.(\ref{top1}).
This behavior is well seen in Fig. \ref{Grenewalb05}, dashed curve.

For $\beta > 1-\alpha/2$, when the integral would diverge at the lower limit if the regularizing
exponential were absent, a new, interesting, intermediate regime arises. 
Close to this lower limit we can approximate the first bracket in the integral in Eq. (\ref{IntegrA}) by unity, and change the variable of integration to $\xi = x^2/4 K_\alpha t^\alpha a^\alpha$ to obtain
\begin{equation}
p(x,t) \simeq
\frac{\left(4K_{\alpha}\right)^{\frac{\beta-1}{\alpha}}}{\alpha\Gamma(\beta)\Gamma(1-\beta)\sqrt{\pi}}t^{\beta-1}|x|^{-1-\frac{2\beta}{\alpha}+\frac{2}{\alpha}}
\int_{\frac{x^2}{4 K_\alpha t^\alpha}}^\infty \xi^{\frac{\beta -1}{\alpha}-\frac{1}{2}} e^{-\xi} d \xi,
\end{equation}
where the integral now represents an upper incomplete Gamma function, which for the intermediate asymptotic domain $x^2 \ll 4 K_\alpha t^\alpha$ can be
approximated by a constant $\Gamma\left(-\frac{1+\beta}{\alpha}-\frac{1}{2}\right)$, so that
\begin{equation}
p(x,t) \simeq
\frac{\left(4K_{\alpha}\right)^{\frac{\beta-1}{\alpha}}\Gamma\left(-\frac{1+\beta}{\alpha}-\frac{1}{2}\right)}{\alpha\Gamma(\beta)\Gamma(1-\beta)\sqrt{\pi}}t^{\beta-1}|x|^{-1-\frac{2\beta}{\alpha}+\frac{2}{\alpha}}
\label{inter1}
\end{equation}
This intermediate asymptotics merges with the top of the distribution 
given by  Eq.(\ref{top}) at $|x| \sim \sqrt{4 K_\alpha \tau_0^\alpha}$, and therefore stretches over the domain $\sqrt{4 K_\alpha \tau_0^\alpha} \ll |x| \ll \sqrt{4 K_\alpha t^\alpha}$
which at large times gets very large. Omitting all prefactors we get
\begin{equation}\label{epdf0b1}
p(x,t)\propto t^{\beta-1}\left|x\right|^{-1-\frac{2\beta}{\alpha}+\frac{2}{\alpha}}
\end{equation}
which can be put into a scaling form $p(x,t) = t^{-\gamma} f(|x|/t^\gamma)$ with $\gamma = \alpha/2$, and $f(z) = z^{\frac{2(1-\beta)}{\alpha}-1}$. 
 
This intermediate asymptotics is seen at long times ($t=1000$) in two further curves depicted in Fig. \ref{Grenewalb05}: for $\alpha= 0.5$ and $\beta=0.9$, where it is very narrow, and for $\alpha =3$ and $\beta = 0.5$,
where it stretches over the whole $x$-domain depicted. For $\alpha=1$, $\frac12<\beta<1$ this expression turns to 
\begin{equation}
p(x,t)\sim t^{\beta-1}\left|x\right|^{1-2\beta}
\end{equation}
which is again in agreement with \cite{NagarGupta}.

\section{MSD and PDF with power-law resetting, $\beta>1$}

\subsubsection{Mean squared displacement} 

Plugging Eqs. (\ref{psireset}) and (\ref{phig1}) into Eq. (\ref{x21}), we get
\begin{equation}\label{xx}
\left\langle x^2(t)\right\rangle = 2K_{\alpha}\tau_0^{\alpha}\left(\beta-1\right)\int_0^{t/\tau_0} d\tau \tau^{\alpha}\left(1+\tau\right)^{-\beta} .
\end{equation}
The integral may be presented in terms of the hypergeometric function
\begin{equation}\label{E2}
\int_0^{t/\tau_0}d\tau\tau^{\alpha}\left(1+\tau\right)^{-\beta}=\frac{\left(t/\tau_0\right)^{1+\alpha}}{1+\alpha}\, _2
F_1\left(1+\alpha,\beta,2+\alpha,-\frac{t}{\tau_0}\right) .
\end{equation}
To get the power-law asymptotics of this integral we use the Affair transformations changing the function 
with the last argument equal to $z$ into a function whose last argument is $z/(z-1)$. There are two variants
of such transformations for $t/\tau_0\to\infty$ \cite{as}
\begin{eqnarray}\label{E3}
\,_2 F_1\left(1+\alpha,\beta,2+\alpha,-\frac{t}{\tau_0}\right) &=& \left(1+ \frac{t}{\tau_0} \right)^{-\beta}\;_2 F_1\left(\beta,1,2+\alpha,\frac{t}{t+\tau_0}\right)\\
_2 F_1\left(1+\alpha,\beta,2+\alpha,-\frac{t}{\tau_0}\right) &=& \left(1+\frac{t}{\tau_0}\right)^{-1-\alpha}\;_2
\,F_1\left(1+\alpha,2+\alpha-\beta,2+\alpha,\frac{t}{t+\tau_0}\right) , \label{E4}
\end{eqnarray}
which, as we proceed to show, are useful in different domains of parameters. For $t \to \infty$ the last argument of the hypergeometric functions on the r.h.s.
of Eqs.(\ref{E3}) and (\ref{E4}) tends to unity, and the corresponding asymptotic values of these functions can be evaluated by applying the Gauss's theorem:
\begin{equation}\label{E5}
_2 F_1\left(a,b,c,1\right)=\frac{\Gamma(c)\Gamma(c-a-b)}{\Gamma(c-a)\Gamma(c-b)}
\end{equation}
valid for 
\begin{equation}\label{E6}
\mathrm{Re}(c)>\mathrm{Re}(a+b).
\end{equation}
Using the transformation Eq.(\ref{E3}) and Eq.~(\ref{E5}) we get for $\beta<\alpha+1$:
\begin{equation}\label{E7}
_2 F_1\left(\beta,1,2+\alpha,1\right)=\frac{1+\alpha}{1+\alpha-\beta}
\end{equation}
and using Eq.(\ref{E4}) and Eq.(\ref{E5}) for $\beta>\alpha+1$ we obtain:
\begin{equation}\label{E8}
_2 F_1\left(1+\alpha,2+\alpha-\beta,2+\alpha,1\right)=\frac{\Gamma\left(2+\alpha\right)\Gamma\left(\beta-\alpha-1\right)}{\Gamma(\beta)} .
\end{equation}
Using the corresponding asymptotic forms in Eq.~(\ref{E2}), and substituting Eq.~(\ref{E2}) into
Eq.~(\ref{xx}), one gets Eq.~(\ref{bla1}) below for $\beta<\alpha+1$ and Eq.~(\ref{ba1}) for $\beta>\alpha+1$.
Thus for $\beta<\alpha+1$ in the long time limit $t\gg\tau_0$ the MSD follows
\begin{equation}\label{bla1}
\left\langle x^2(t)\right\rangle = \frac{2K_{\alpha}\tau_0^{\beta-1}\left(\beta-1\right)}{\alpha-\beta+1}t^{\alpha+1-\beta} .
\end{equation}
The result is presented at the Fig. \ref{GR2exprenewal} as a gray line. In the case of ordinary Brownian motion $\alpha = 1$ and $\beta<2$ the motion appears to be subdiffusive
\begin{equation}
\left\langle x^2(t)\right\rangle  = 2K_{1}\tau_0^{\beta-1}t^{2-\beta}\frac{\beta-1}{2-\beta}\,.
\label{x2b12obm}
\end{equation}
For $\beta>\alpha+1$ the MSD stagnates:
\begin{equation}\label{ba1}
\left\langle x^2(t)\right\rangle =
\frac{2K_{\alpha}\tau_0^{\alpha}\left(\beta-1\right)}{\alpha+1}\cdot\frac{\Gamma\left(2+\alpha\right)\Gamma\left(\beta-\alpha-1\right)}{\Gamma(\beta)}.
\end{equation}
In the case of ordinary Brownian motion, $\alpha=1$, the MSD stagnates and attains the value \begin{equation}
\left\langle x^2(t)\right\rangle  = \frac{2K_{1}\tau_0}{\beta-2}\,.
\label{x2b2obm}
\end{equation}
The corresponding numerical results are presented at Fig. \ref{GR2exprenewal} with red ($\beta=2.5$) and orange ($\beta=10$) lines, respectively.

\subsubsection{Probability density function} 

The probability density function for $\beta>1$ may be obtained by inserting Eqs. (\ref{p0sbm}), (\ref{psireset}) and
(\ref{phig1}) into Eq. (\ref{eqprob1})
\begin{equation}
p(x,t) \simeq \frac{\beta-1}{\tau_0}\frac{1}{\sqrt{4\pi K_{\alpha}}}\int_0^t
dt^{\prime}\left(1+\frac{t-t^{\prime}}{\tau_0}\right)^{-\beta}\left(t-t^{\prime}\right)^{-\alpha/2}\exp\left(-\frac{x^2}{4K_{\alpha}\left(t-t^{\prime}\right)^{\alpha}}\right).
\end{equation}
\color{black}
We note that assuming the rate of events constant is possible only for $t \gg \tau_0$, and therefore all the results below are valid only for $t \gg \tau_0$, under
which condition also the proper normalization is guaranteed. This implies that if $p(x,t)$ possesses singularities, these must be integrable.
Putting $x=0$ one readily infers that for $\alpha < 2$ the integral converges, so that $p(x,t)$ is finite at the origin and develops a flat top, and moreover
$p(0,t)$ tends to a finite limit for $t \to \infty$. On the other hand, for $\alpha > 2$ the PDF at the origin diverges.

Changing  to the dimensionless variable of integration $z=\frac{x^2}{4K_{\alpha}\left(t-t^{\prime}\right)^{\alpha}}$ one gets
\begin{equation}
p(x,t) \simeq \frac{\beta-1}{\tau_0 \alpha \sqrt{4\pi K_{\alpha}}}\left(\frac{x^2}{4K_{\alpha}}\right)^{-\frac{1}{2}+\frac{1}{\alpha}}
\int_{\frac{x^2}{4K_{\alpha} t^{\alpha}}}^{\infty}  \left[1 + \frac{1}{\tau_0} \left(\frac{x^2}{4K_{\alpha}} \right)^{\frac{1}{\alpha}} z^{-\frac{1}{\alpha}} \right]^{-\beta} z^{-\frac{1}{2}-\frac{1}{\alpha}} e^{-z} dz.
\label{pdf_Gen}
\end{equation}
Since for large $z$ the integrand is strongly suppressed by the decaying exponential, the unity in square brackets can be 
safely neglected for $x^2 \gg 4 K_\alpha \tau_0^\alpha$, and the integral can be approximated by 
\begin{equation}\label{pxtint}
p(x,t) \simeq \frac{\left(\beta-1\right)\tau_0^{\beta-1}}{\alpha\sqrt{4\pi K_{\alpha}}}\left(\frac{x^2}{4K_{\alpha}}\right)^{-\frac{1}{2} + \frac{1}{\alpha}- \frac{\beta}{\alpha}  }
\int_{\frac{x^2}{4K_{\alpha}t^{\alpha}}}^{\infty} dz e^{-z} z^{-\frac{1}{2}-\frac{1}{\alpha}+ \frac{\beta}{\alpha}} .
\end{equation}
so that
\begin{equation}
p(x,t) \simeq \frac{\left(\beta-1\right)\tau_0^{\beta-1}}{\alpha\sqrt{4\pi
K_{\alpha}}}\left(\frac{x^2}{4K_{\alpha}}\right)^{-\frac{1}{2}  + \frac{1}{\alpha}- \frac{\beta}{\alpha} } \Gamma\left(\frac{1}{2} + \frac{\beta}{\alpha} - \frac{1}{\alpha} ,
\frac{x^2}{4K_{\alpha}t^{\alpha}}\right) .
\label{pdfb15}
\end{equation}

Eq. (\ref{pdf_Gen}) also allows for determining the type of singularity at the origin for $\alpha > 2$. For $x^2 \ll 4 K_\alpha \tau_0^\alpha$ we can neglect the second 
term in the square brackets in Eq. (\ref{pdf_Gen})
\begin{equation}
p(x,t)=\frac{\beta-1}{\tau_0 \alpha \sqrt{4\pi K_{\alpha}}}\left(\frac{x^2}{4K_{\alpha}}\right)^{-\frac{1}{2}+\frac{1}{\alpha}}
\int_{\frac{x^2}{4K_{\alpha} t^{\alpha}}}^{\infty} z^{-\frac{1}{2}-\frac{1}{\alpha}} e^{-z} dz.
\end{equation}
and then take the limit $x \to 0$ in the lower bound of integration, so that the integral converges to a Gamma function  $\Gamma(1/2 - 1/\alpha)$ of a 
non-negative argument. Therefore for $\alpha > 2$ one gets $p(x,t) \sim |x|^{2/\alpha -1}$ showing an integrable singularity. 

\begin{figure}[htbp]
  \centerline{
\includegraphics[width=0.7\columnwidth]{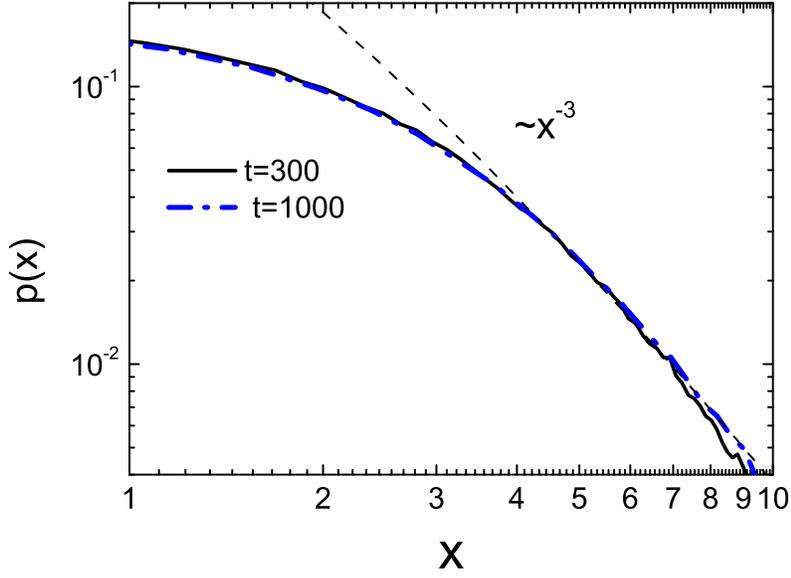}}
\caption{PDF for SBM with power-law resetting for $t=300$ and 1000 showing the steady state. The black dashed line corresponds to fitting with
$p(x,t)\simeq x^{-3}$ (Eq.~\ref{pba}). Parameters: $N=10^6$, $\alpha=0.5$, $\beta=1.5$.}
\label{Grenewalb15a05}
\end{figure}

Let us return to our Eq. (\ref{pdfb15}) and discuss the behavior of the PDF in the intermediate asymptotic regime $\sqrt{4 K_\alpha \tau_0^\alpha} \ll x \ll \sqrt{4 K_\alpha t^\alpha}$
which, for $t$ large, stretches over the whole relevant domain of $x$. We note that for small values of the second argument $y \to \infty$ the incomplete Gamma function $\Gamma(a,y)$ 
tends to $\Gamma(a)$ while for large values of $y$ it possesses the asymptotics $\Gamma(a,y) \simeq y^{a-1} e^{-y}$. 
Therefore at long times $t^{\alpha}\gg x^2/(4K_{\alpha})$ the PDF, Eq. (\ref{pdfb15}), for $x$ fixed tends to the steady state
\begin{equation}\label{pba}
p(x,t) \sim |x|^{-1-2\beta/\alpha+2/\alpha} .
\end{equation}
The results of numerical simulation of PDF (Fig. \ref{Grenewalb15a05}) confirm this finding. The scaling $p(x,t)\simeq
x^{-3}$ (for $\beta=1.5$, $\alpha=0.5$) nicely fits the obtained results at large values of $x$. For $\alpha=1$ the PDF scales as $p(x,t)\sim
x^{1-2\beta}$ as obtained in \cite{NagarGupta}. 

It is interesting to note that the PDF in its bulk tends to the steady state at long time for the whole
range of $\beta>1$, while the MSD stagnates only for $\beta>1+\alpha$ but grows in the course of time when $1<\beta<1+\alpha$, see Eqs. (\ref{bla1}) and (\ref{ba1}), which fact stresses the
absence of the overall scaling. 
To explain this phenomenon we return to Eq. (\ref{pdfb15}) and note that for $x \gg \sqrt{4 K_\alpha t^\alpha}$ the power-law
as given by Eq. (\ref{pba}) has a Gaussian cutoff (due to the asymptotics of incomplete Gamma function discussed above). In the case when the second moment of the approximate
PDF in the center as given by Eq. (\ref{pba}) converges, $\int_{-\infty}^\infty x^2 p(x,t) dx < \infty$, i.e. for $\beta > 1 + \alpha$, this cutoff does not play any role provided the central part is broad enough, i.e. for long times. 
If the corresponding integral diverges, the second moment is governed by the position of the cutoff at $x \sim \sqrt{4 K_\alpha t^\alpha}$, and behaves exactly as predicted by Eq. (\ref{bla1}).

\color{black}

\section{Mean first passage time under power-law resetting}

The MFPT for resetting with power-law waiting time distribution can be obtained in full analogy to the case of the exponential resetting.
In order to calculate the first-passage time for the power-law resetting we use Eq.~(\ref{fpt}) with Eqs.~(\ref{X0}) and (\ref{R0}), with survival probability of
target, Eq.~(\ref{Phit}), the waiting time distribution between resetting events, Eq.~(\ref{pdfpow}), and the resetting survival probability,
Eq.~(\ref{psireset}). The expression for $X_0$ reads:
\begin{equation}
X_0=\frac{2}{\sqrt{\pi}}\int_0^{\infty}dt\left(1+\frac{t}{\tau_0}\right)^{-\beta}\int_0^{\frac{x_0}{2\sqrt{K_{\alpha}t^{\alpha}}}}dae^{-a^2} .
\end{equation}
By changing the order of integrations we arrive at the expression
\begin{equation}
X_0=\frac{2}{\sqrt{\pi}}\int_0^{\infty}dae^{-a^2}\int_0^{\left(\frac{x_0^2}{4K_{\alpha}a^2}\right)^{\frac{1}{\alpha}}}\left(1+\frac{t}{\tau_0}\right)^{-\beta}dt .
\end{equation}
Now the inner integration may be performed explicitly to yields for the numerator in Eq.~(\ref{fpt})
\begin{equation}
X_0=\frac{2\tau_0}{\sqrt{\pi}\left(1-\beta\right)}\int_0^{\infty}dae^{-a^2}\left(1+\left(\frac{A}{a^2}\right)^{\frac{1}{\alpha}}\right)^{1-\beta}-\frac{\tau_0}{1-\beta}
\end{equation}
with 
\begin{equation}\label{A}
A=\frac{1}{\tau_0}\left(\frac{x_0^2}{4K_{\alpha}}\right)^{\frac{1}{\alpha}}.
\end{equation}
On the other hand, denominator in Eq.~(\ref{fpt}) reads 
\begin{equation}
1-R_0=\frac{2}{\sqrt{\pi}}\int_0^{\infty}dae^{-a^2}\left(1+\left(\frac{A}{a^2}\right)^{\frac{1}{\alpha}}\right)^{-\beta} .
\end{equation}
Changing the variables of integration in both expressions to $y=a^2$ we get
\begin{eqnarray}
X_0 &=&
\frac{\tau_0}{\sqrt{\pi}\left(1-\beta\right)}\int_0^{\infty}dyy^{-\frac{1}{2}}\left(1+\left(\frac{A}{y}\right)^{\frac{1}{\alpha}}\right)^{1-\beta}e^{-y}-\frac{\tau_0}{1-\beta}\\
1-R_0 &=& \frac{1}{\sqrt{\pi}}\int_0^{\infty}dyy^{-\frac{1}{2}}e^{-y}\left(1+\left(\frac{A}{y}\right)^{\frac{1}{\alpha}}\right)^{-\beta}\label{G6}
\end{eqnarray}
Let us evaluate Eq.~(\ref{G6}). 
The integral can be roughly estimated by splitting the integration domain into two parts (at the point $A$) and neglecting subleading terms:
\begin{equation}
\int_0^{\infty}dyy^{-\frac12}e^{-y}\left(1+Ay^{-\frac{1}{a}}\right)^{-\beta} \simeq
\int_0^{A}dye^{-y}A^{-\beta}y^{\frac{\beta}{a}-\frac12}+\int_A^{\infty}dyy^{-\frac12}e^{-y} .
\end{equation}
In the first integral we have neglected unity compared to $Ay^{-\frac{1}{a}}$ and in the second one we neglect 
$Ay^{-\frac{1}{a}}$ compared to 1. Both integrals are now incomplete Gamma functions. $X_0$ can be calculated analogously. 
The final result reads 
\begin{equation}\label{taufull}\tau \simeq \frac{\tau_0}{\beta-1}\times\frac{\sqrt{\pi}\rm
erf(\sqrt{A})-A^{1-\beta}\gamma\left(A,\frac{\beta-1}{\alpha}+\frac12\right)}{A^{-\beta}\gamma\left(A,\frac{\beta}{\alpha}+\frac12\right)+\sqrt{\pi}\rm
erfc(\sqrt{A})}.
\end{equation}
For $A = \frac{1}{\tau_0}\left(\frac{x_0^2}{4K_{\alpha}}\right)^{\frac{1}{\alpha}}\to\infty$ corresponding to the target located far from origin
the MFPT tends to
\begin{equation}
\label{tau}
\tau=\frac{\sqrt{\pi}\tau_0^{1-\beta}}{\left(\beta-1\right)\Gamma\left(\frac{\beta}{\alpha}+\frac12\right)}\left(\frac{x_0^2}{4K_{\alpha}}\right)^{\frac{\beta}{\alpha}} .
\end{equation}

\begin{figure}[htbp]
  \centerline{
\includegraphics[width=0.7\columnwidth]{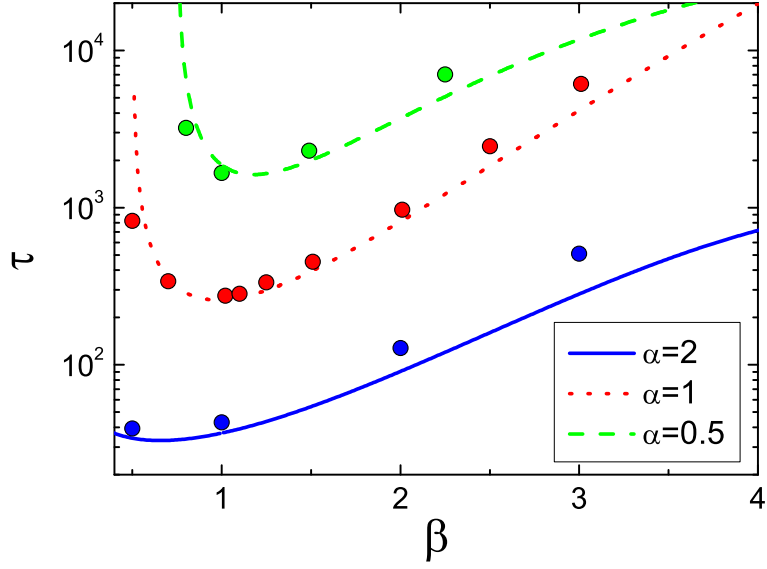}}
\caption{Mean FPT as function of $\beta$ for $x_0=10$. 
The results of numerical simulation are depicted by circles. The analytical results, Eq.~(\ref{taufull}), are given by lines.}
\label{GTpow}
\end{figure} 
The results of numerical simulations together with the predictions of Eq.~(\ref{taufull}) are shown in Fig. \ref{GTpow}.
One can see that for distant targets ($x_0=10$) the search
becomes more efficient in the case of superdiffusion in a large domain of $\beta$.

\section{Conclusions}

In the present work we study analytically and numerically the properties of scaled Brownian motion with time-dependent diffusion coefficient $D(t)\sim
t^{\alpha-1}$ interrupted by random resetting to the origin. The resetting process is a renewal one with the PDF of waiting times between the renewal events being either exponential or 
a power law with $\psi(t) \sim t^{-1-\beta}$. 
In the present work we concentrate on the situation, where the diffusion coefficient is also reset, and
the displacement process is rejuvenated, so that the whole process is a renewal one. Erasing the memory in the transport process under reseting events
is of large importance for the overall behavior, making its dynamics different from the case when the memory in the transport process is 
retained (i.e. the diffusion coefficient is not reset), as discussed in the other work of these series. 

The main results of the present work are as follows: For exponential resetting and power-law resetting with $\beta>1+\alpha$ the  MSD at long times stagnates.
\color{black} For exponential resetting this fact was already proved for a general class of anomalous diffusion processes \cite{Maso1}. \color{black}
For $\beta<1$ the time dependence of the MSD remains the same as in the case of free scaled Brownian motion, albeit with different prefactors. 
In the intermediate domain $1<\beta<1+\alpha$ we obtain $\left\langle x^2\right\rangle\sim t^{1+\alpha-\beta}$, so that the behavior of the MSD is defined by the interplay of the parameters $\alpha$ and $\beta$. 

In the case of the exponential resetting the PDF tends to a steady state with stretched or squeezed exponential tail $p(x,t)\simeq \exp\left(-\gamma
|x|^{\frac{2}{\alpha+1}}\right)$. For the power-law resetting with $\beta>1$ the PDF also attains a time-independent form, now $p(x,t)\sim x^{-1-\frac{2\beta}{\alpha}+\frac{2}{\alpha}}$. 
It is interesting to note that for $\beta>1+\alpha$ both the MSD and the PDF tend to the stationary state, while for $1<\beta<1+\alpha$ only the PDF in the bulk is stationary but the
MSD grows continuously with time. For $\beta<1$ the behavior of the PDF depends on the relation between the
exponents $\beta$ and $\alpha$. For $\beta > 1 - \alpha/2$ the $x$-dependence of the PDF for $\sqrt{4 K_\alpha \tau_0^\alpha} \ll |x| \ll \sqrt{4 K_\alpha t^\alpha}$ is the same as in the previous case,
but now the time-dependence also appears: $p(x,t)\sim t^{\beta-1}\left|x\right|^{-1-\frac{2\beta}{\alpha}+\frac{2}{\alpha}}$. For long times this intermediate domain covers
practically the whole bulk of the distribution. For $\beta < 1 - \alpha/2$ the PDF in the center of the distribution is flat, with a Gaussian tail at $x \gg \sqrt{4 K_\alpha t^\alpha}$. 
Below we present the MSD and PDF in the form of the Table \ref{TableI}.  The results for the time dependence of the MSD for power law resetting with $\beta < 1$ coincide with those of Ref. \cite{Maso1}.

\begin{table}[h!]
\caption{MSD and PDF for the renewal power-law resetting \label{TableI}}
\centering
\begin{tabular}{ | l | l | l | l |  l | }
\hline
 & $0<\beta<1-\alpha/2$ & $1-\alpha/2 < \beta < 1$ & $1<\beta<1+\alpha$ & $\beta>1+\alpha$ \\ \hline
MSD & $\sim t^{\alpha}$ & $\sim t^{\alpha}$ & $\sim t^{\alpha+1-\beta}$ & stagnates \\
PDF &flat top, Gaussian tail & $\sim t^{\beta-1}\left|x\right|^{-1-2\beta/\alpha+2/\alpha}$ & $\sim \left|x\right|^{-1-2\beta/\alpha+2/\alpha}$ & $\sim
\left|x\right|^{-1-2\beta/\alpha+2/\alpha}$ \\
\hline
\end{tabular}
\end{table}

The overall renewal nature of the whole process allowed us also to calculate the mean first passage time to a target.
This MFPT is investigated as a function of parameters of the model for the corresponding cases of Poissonian and power-law resetting. As it is generally the case, resetting makes the search of the target much more effective. 
There always exists an optimal resetting strategy minimizing the MFPT. The subdiffusive search is favorable at large resetting rates and for remote targets. The superdiffusion is more efficient at small resetting rates and for target locations close to the starting point. 

\color{black}
The results for the renewal resetting scheme for SBM should be confronted with the ones for the situation when the transport process is not rejuvenated under resetting, and the whole process
is non-renewal, as discussed in detail in the other work of this series \cite{Anna0}. The behavior observed in this process significantly differs from the results discussed above. 
Here the MSD in the case of exponential resetting does not stagnate for $\alpha \neq 1$ and shows the behavior $\langle x^2 \rangle \sim t^{\alpha -1}$.  
The time dependence of the MSD for power-law resetting is summarized in Table \ref{TableII}. Thus for slowly decaying waiting time PDFs with $\beta < 1$ this MSD follows  $\langle x^2 \rangle \sim t^{\alpha}$,
like in the free SBM, while for $\beta > 1$ the growth gets slower, and for rapidly decaying PDFs with $\beta > 2$ the time dependence of the MSD is the same as in the exponential case, namely 
$\langle x^2 \rangle \sim t^{\alpha-1}$. This implies that for subdiffusive SBM with $\alpha < 1$ the particle gets localized at the origin. 

\begin{table}[h]
\caption{MSD and PDF for the non-renewal power-law resetting \label{TableII}}
\centering
\begin{tabular}{ | l | l | l |  l | l | }
\hline
 & $0<\beta<1/2$ & $1/2<\beta<1$ & $1<\beta<2$ & $\beta>2$ \\ \hline
MSD & $\sim t^{\alpha}$ & $\sim t^{\alpha}$ & $\sim t^{\alpha+1-\beta}$ & $\sim t^{\alpha-1}$ \\
PDF & $\begin{array}{cc}
       \mbox{flat top, Gaussian tail}
      \end{array}$
 & $\sim t^{\alpha(\beta-1)}\left|x\right|^{1-2\beta}$ & $\sim t^{(1-\beta)(1-\alpha)} \left|x\right|^{1-2\beta}$ & $\sim t^{(1-\beta)(1-\alpha)}
\left|x\right|^{1-2\beta}$ \\
\hline
\end{tabular}
\end{table}

In contrast with the renewal case, the PDF of the particle's position for non-renewal resetting with exponential waiting time always shows simple 
two-sided exponential (Laplace) shape and is non-stationary (cf. with stationary stretched or squeezed exponential forms discussed above). 
In the case of power-law resetting time PDF with very slow decay ($\beta < 1/2$) the PDF of positions does not show any universal scaling in the body and possesses Gaussian tails. 
In all other cases it tends to universal forms which are different for $1/2 < \beta < 1$, and for $\beta > 1$. These forms are however different from the ones obtained in the renewal case. 
The main physical consequence of our discussion is that it shows that erasing or retaining the memory in transport process is crucial for the features of the overall dynamics. 
\color{black}

\color{black}
Our results have several implications going beyond the standard resetting scheme and displacement process as represented by SBM. 
We note that the PDF of the random process under renewal resetting depends only on the PDF of displacements of the free displacement process
within a single renewal epoch. Moreover, the MSD in such a process depends only on the MSD in a displacement process, and not on other properties of 
this process, such as its PDF or correlations between the increments.

These statements mean that under fully renewal setup, the results for PDF for \textit{any} Gaussian displacement process will be the same as for the SBM.
Moreover, the corresponding formulas can be applied to \textit{increments} of any Gaussian process with  \textit{stationary increments} sampled at random times following a renewal process.
Physically, we here consider our resetting not as a physical return to the origin but as declaring of the actual particle's position at the end of a renewal epoch as a new origin. 
As an example, one can consider increments of a fractional Brownian motion \cite{LimSBM}, a non-Markovian process with stationary increments with nontrivial memory.
If this dependence is $\langle [x(t+\Delta t)-x(t)]^2 \rangle \sim \Delta t^\alpha$, the PDF of displacements and the MSD in the time intervals 
defined by the sampling times with power-law distribution of periods between them will follow from Table \ref{TableI}.

We moreover note that in the subdiffusive case the SBM can be considered as a mean-field (Gaussian) approximation for the CTRW model
\cite{ThielSok} with a power-law waiting time probability density function, which, like the SBM, is a process with non-stationary increments. In SBM this non-stationarity
is modeled via the explicit time dependence of the diffusion coefficient, while the CTRW, being of the renewal class, lacks explicit time
dependences of its parameters. On the other hand, SBM is a Markovian process, while CTRW is a non-Markovian (semi-Markovian) one. Nevertheless, aging
properties of both processes are very similar, and so should be the behaviors of the MSD under two types of resetting. The probability density functions of the processes 
however differ. 

The analogies above cannot be generalized to the first hitting times which depend on other properties of the motion in a single renewal epoch (multipoint probability
densities) than the singe-time PDF.

\color{black}


\section{Acknowledgements}

A.S.B. thanks Carlos Meji\'a-Monasterio, Arnab Pal and Shlomi Reuveni for fruitful discussions. AVC acknowledges the financial support by the Deutsche Forschungsgemeinschaft within the project ME1535/6-1.\\

\end{document}